\documentclass[10pt,journal,cspaper,compsoc]{journeevisu}

\usepackage[pdftex]{graphicx} 

\usepackage{t1enc}
\usepackage{url}
\usepackage{cite}
\usepackage{hyperref}
\usepackage{flushend}

\usepackage{orcidlink}

\TitreVisu{Situated Visualization in Motion for Swimming}{Visualisation Localisée en Mouvement pour la Natation}

\ShortBib{Lijie Yao \MakeLowercase{\textit{et al.}}: Situated Visualization in Motion for Swimming}

\author{Lijie Yao \orcidlink{0000-0002-4208-5140}, Anastasia Bezerianos \orcidlink{0000-0002-7142-2548}, Romain Vuillemot \orcidlink{0000-0003-1447-6926}, and Petra Isenberg \orcidlink{0000-0002-2948-6417}
\IEEEcompsocitemizethanks{
\IEEEcompsocthanksitem Lijie Yao: Universit{\'e} Paris-Saclay, CNRS, Inria, LISN, Orsay, France;
  \protect E-mail: yaolijie0219@gmail.com.
\IEEEcompsocthanksitem Anastasia Bezerianos: Universit{\'e} Paris-Saclay, CNRS, Inria, LISN, Orsay, France;
  \protect E-mail: anastasia.bezerianos@universite-paris-saclay.fr.
\IEEEcompsocthanksitem Romain Vuillemot: Universit{\'e} de Lyon, {\'E}cole Centrale de Lyon, CNRS, UMR5205, LIRIS, France;
  \protect E-mail: romain.vuillemot@ec-lyon.fr.
\IEEEcompsocthanksitem Petra Isenberg: Universit{\'e} Paris-Saclay, CNRS, Inria, LISN, Orsay, France;
    \protect E-mail: petra.isenberg@inria.fr.
}
}

\IEEEcompsoctitleabstractindextext{

\begin{abstract}
  Competitive sports coverage increasingly includes information on athlete or team statistics and records. Sports video coverage has traditionally embedded representations of this data in fixed locations on the screen; but more recently also attached representations to athletes or other targets in motion. 
  These publicly used representations so far have been rather simple and systematic investigations of the research space of embedded visualizations in motion is still missing. Here we report on our preliminary research in the domain of professional and amateur swimming. We analyzed how  visualizations are currently added to the coverage of Olympics swimming competitions and then plan to derive a design space for embedded data representations for swimming competitions. We are currently conducting a crowdsourced survey to explore which kind of swimming-related data general audiences are interested in, in order to identify opportunities for additional visualizations to be added to swimming competition coverage. 
\end{abstract}

}

\begin{document}

\maketitle

\section{Introduction}
We define visualizations in motion as visual data representations that are used in contexts that exhibit relative motion between a viewer and an entire visualization \cite{yao2020situated}.  There are already meany scenarios that involve visualizations in motion such as sports analytics, video games, wearable devices, and data physicalizations. However, the relative movement between viewers and visualizations in scenarios above has not received much attention in the visualization research community and we know very little about how effectively people can read these visualizations.

To explore this research space we first focused on sports analytics as an application scenario in which embedded moving visualizations are particularly promising. In many sports athletes are in motion and stationary viewers are interested in seeing a data such as statistics, past records, trajectories, etc. We began our work by focusing on one specific sport, swimming, for the following reasons:
Like other sports it has rich dynamic data and we have observed an increase of real-time moving graphics during competitions (e.g., record lines), sometimes attached to athletes (e.g., country flags or speed labels). This indicates that audiences may be open to the addition of more complex moving data representations. Moreover, it is a sport where the movement of athletes is constrained in lanes, making it an appealing starting point for our investigation into how viewers perceive and track this data.

To better understand the visualization opportunities in this context, we first analyzed current visualizations embedded in swimming competitions. We next plan to understand what other data we could visualize. Currently, we are conducting a survey to explore the information interests of general audiences when watching a swimming competition. Our next phase is to develop a prototype with embedded moving visualizations, to show the data of interest to  general audiences, and to evaluate the efficiency of \emph{visualization in motion} in an actual concrete application context.

\vspace{-10pt}
\section{Related Work}
Our research is closely related to the topic of situated and embedded visualization \cite{Willett:2017:EDR} that represent information close to a data referent. Specifically, we focus on embedded visualizations that are in motion. 
Although there is little discussion in the visualization community about \emph{relative} motion between the entire visualization and the viewers, visualization research involving movement inside the visualization is extensive. For example, animation is frequently used for smooth transition between different data points \cite{GraphDiaries:2014, StaggeringAnimation:2014}, or to morph between different representations \cite{3DTreemaps:2005, StagedAnimation:2007, ConeTrees:1991}. In contrast, in our work we focus on the potential impact and application scenarios when the entire visualization is moving rather than its individual components.

A large number of visual analytics tools have been developed  for sports visualization. Examples include tools for table tennis \cite{TabletennisVis3D, Wang:2021:TableTenis, Wu:2021:TableTennisVR}, soccer \cite{Chen:2020:ARSoccer, perin:2013:soccerdata, SoccerTeamSportAnalysis:2017}, basketball \cite{Lin:2021:ARVisforBasketballTraining, Bai:2016:BasketballAnalysis}, badminton \cite{Pingali:2001:BadmintonAnalysis, BadmintonVis3D}, and tennis \cite{TenisVis}. In contrast to entire moving visualizations, these tools consistently placed visualizations at specific fixed locations: for example, in VisCommentator \cite{Chen:2021:VisCommentator}, visualizations are shown when video playback is paused, rather than embedded in the video stream. Besides, the target users for most previous visualization tools were experts, such as coaches and/or athletes; whereas, we target general audiences.

\begin{figure}
    \centering
    \includegraphics[height=2.2cm]{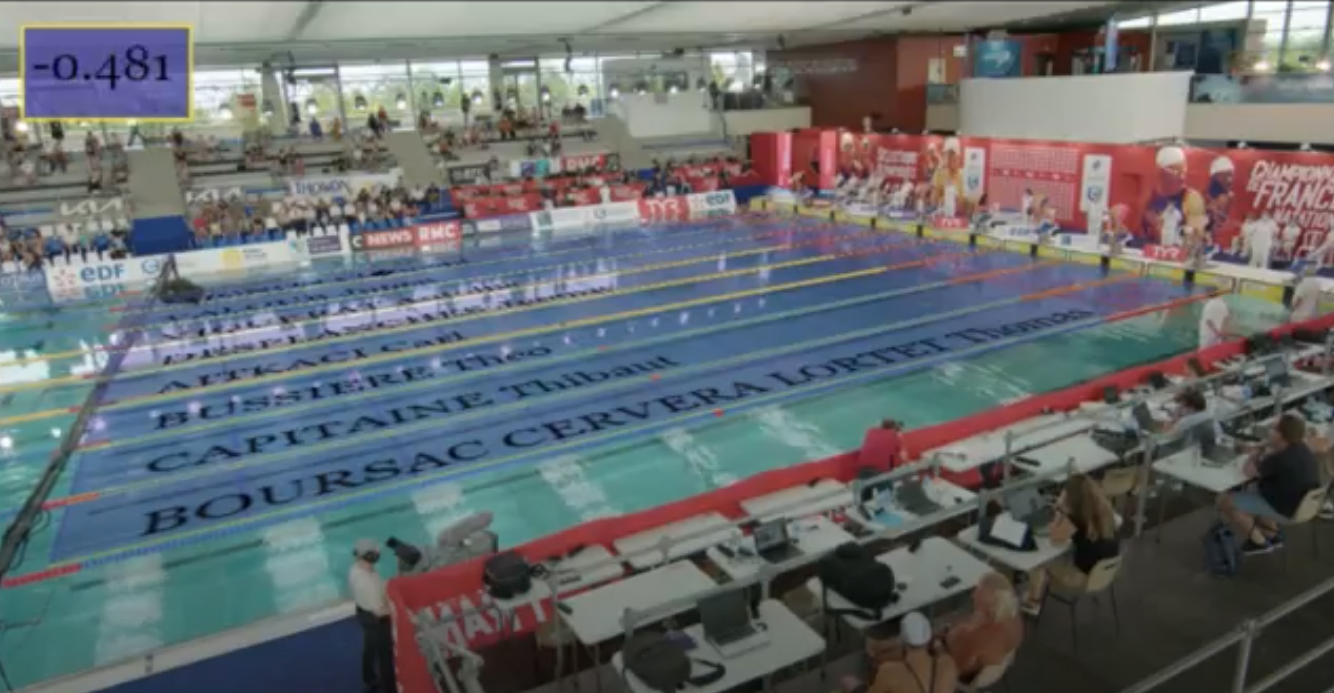}\hfill
    \includegraphics[height=2.2cm]{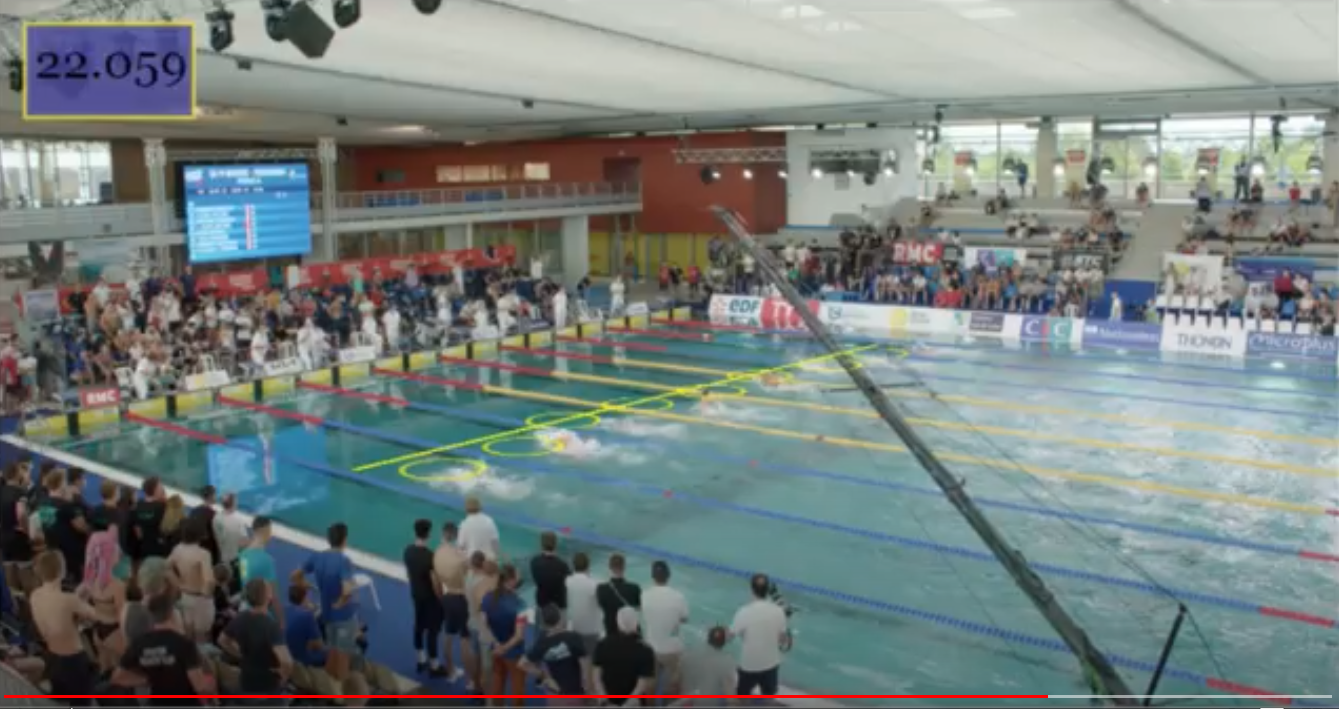}
    \caption{Examples of embedded visualizations for swimming. Left: Stationary swimmers' names in each lane. Right: Swimmers' current position circles and record line moving with the swimmers' movement. Both images have a stationary timer in the top left corner.\\
    \footnotesize Image credits: We analyzed the videos from the 2020 Olympics. Since we did not obtain permission to use screenshots of these videos, we use our own recordings of other competitions. }
    \label{fig:MovingLine}
\end{figure}


\section{Swimming Visualization Analysis}
As our research focuses on \emph{visualization in motion} in swimming competitions, we  investigated a) which data is displayed in current competitions, b) using which representations, c) where on the screen, and d) under which camera positions and perspectives. 
We analyzed all 16 available swimming videos of the 2020 Tokyo Olympics from the official Olympic Aquatics YouTube  channel \cite{OlympicAquatic}. We picked all 16 available videos so that they would cover all 4 strokes and all 5 race lengths (50m -- 400m, including 4 \texttimes\ 100m medlay). Since we were interested in scenarios where the entire visualizations moved, we only focused on the swimming process. That is, we do not discuss any visualizations shown outside of the swimming activity itself such as when athletes stand before the race or during the awards ceremony. We found: 

\noindent\textbf{Data items:} There were three categories of data shown in the broadcast: swimmers': nationalities, names, lane number, current speed, and distance swam; temporal information, and record-related information. 
Temporal information included the time taken from the start of the race to the current time and the elapsed time difference between the current swimmer and a specific record. The records shown were the world record and the Olympic record.

\noindent\textbf{Representations:} We saw three types of representations: symbols, text, and marker lines. Symbols represented nationality using a flag; text represented lane numbers, names, times, speeds, and distances; and line marks represented where a swimmer should be to match a record. Only two visualizations moved in the videos we surveyed: text moved behind the swimmer to show their current speed and the record marker line moved to show an interpolated position of a record (see \autoref{fig:MovingLine}: Right).

\noindent\textbf{Display positions:} Apart from the two moving visualizations mentioned above, the rest of the visualizations remained in a fixed location. They were attached to the corners of the screen or embedded in the swimming pool (see \autoref{fig:MovingLine}: Left).
The top left, bottom left and bottom right corners were the three most frequently used display areas for real-time changing data. When a swimmer turned or a handoff occurred, the current standings and elapsed time appeared in the upper left corner (see \autoref{fig:MovingLine}: Left), the record and the gap to the record appeared in the lower left corner, and the current timing and meta information about the competition appeared in the lower right corner. Once the first three swimmers arrived their names, nationalities, and lanes were also embedded in the pool.

\noindent\textbf{Camera positions and perspectives:} There were two main camera positions: in the air and under water. The camera perspectives were more varied. We saw bird's-eye views (top), side views, and diagonal views. In the diagonal view, swimmers swam along the diagonal of the screen -- from the bottom right to the top left or from the bottom left to the top right corner. According to our statistics on the display time of each view, the diagonal view was visible the longest. When the camera moved or the camera perspective changed, the moving and embedded visualizations were deformed to match the current view. 

\section{Work in Progress: Survey}
After analyzing Olympic swimming coverage, we found that the displayed data items had not changed for over 20 years, and the visualizations used were consistently simple and basic. 
To understand if people's preferences and interest in information on swimming has changed we are now conducting a crowdsourcing survey.

\section{Discussion and Future Work}
While our survey is running we are preparing a prototype for embedding different data visualizations in swimming videos. We will apply our survey findings to our prototype: we will embed the most interesting data next to the swimmers. In the future, we want to evaluate the impact of contextual factors, such as the background or camera perspectives on \emph{visualizations in motion}. We also plan to examine how efficient visualizations in motion are and how much information viewers can perceive in a real application scenario. 

\section{Acknowledgement }
This work was partly supported by the Agence Nationale de la Recherche (ANR), grant number ANR-19-CE33-0012. 
The authors thank all volunteers who participated in our survey.

\bibliographystyle{abbrv}
\bibliography{abstract}

\end{document}